\documentclass[epjST]{svjour}
\usepackage{graphicx}
\usepackage{dcolumn}
\usepackage{amsmath}
\usepackage{amssymb}
\begin{document}
\title{Topological phases of a Kitaev tie}
\author{Alfonso Maiellaro\inst{1}\fnmsep\thanks{\email{amaiellaro@unisa.it}} \and Francesco Romeo\inst{1} \and Roberta Citro\inst{1,2}}
\institute{ Dipartimento di Fisica ''E. R. Caianiello'',
	Universit\`a degli
	Studi di Salerno, Via Giovanni Paolo II, I-84084 Fisciano (SA), Italy \and Spin-CNR, Universit\`a di Salerno, I-84084 Fisciano (SA), Italy }
\abstract{We investigate the topological properties of a Kitaev chain in the shape of a legged-ring, which is here referred to as Kitaev tie. We demonstrate that the Kitaev tie is a frustrated system in which topological properties are determined by the position of the movable bond (the tie knot). We determine the phase diagram of the system as a function of the knot position and chemical potential, also discussing the effects of topological frustration. The stability of the topological Kitaev tie is addressed by a careful analysis of the system free energy.}
\maketitle
\section{Introduction}
\label{intro}
Kitaev first proposed a minimal model \cite{Kitaev} which exhibits the nucleation of Majorana modes (MMs) at the extremal points of a one-dimensional chain of spinless fermions subject to a $p$-wave superconducting pairing. Low-dimensional systems hosting Majorana modes represent a promising playground for the implementation of fault-tolerant quantum computation and, accordingly, Majorana quasi-particles have attracted increasing attention in the scientific literature \cite{fukane2008,mourik12,nadj1,nadj2,beenakker,aguado,mazziotti,Maiellaro1,Maiellaro2}. Interestingly, when periodic boundary conditions are assigned to the Kitaev chain (Kitaev ring) a non-topological system is obtained with no unpaired Majorana fermions at the system edges. A condition of topological frustration can be induced in the Kitaev chain by simply adding an extra hopping between two arbitrarily distant sites. Depending on the selected sites, the original Kitaev chain, which is a topological non-trivial system, can be transformed into a topological-trivial Kitaev ring or into a legged-ring $\varpropto$, whose topological properties cannot be easily anticipated. Starting from a legged-ring and progressively reducing the ring circumference, a physical system equivalent to a Kitaev chain perturbed by a local defect is obtained, the latter system probably falling into the topological non-trivial class. The interesting point is that the topological properties, which are usually quite robust against e.g. the disorder effects, are here strongly perturbed by a minimal modification of the Kitaev Hamiltonian, i.e. the addition of a single bond. From the above arguments, considering the process in which a non-local bond is shifted in order to progressively reduce the circumference of a legged-ring, we might think that a topological phase transition occurs when a critical size of the ring circumference is reached. This simple expectation is not actually realized and trivial and non-trivial phases alternate as the circumference of the legged-ring is reduced. This behavior, which clearly reflects the topological frustration of the system, is the object of the present analysis.\\
The interest towards Kitaev chains with an extra hopping between two arbitrarily distant sites is not a mere theoretical curiosity. Indeed it represents the minimal effective model of a geometry which is realized in looped single-walled carbon nanotubes (SWCNTs)\cite{Jespersen}, which have been used to implement proof-of-principle nanodevices \cite{Refael}. SWCNTs are flexible ballistic conductors with semiconducting or metallic properties \cite{swcn}. Proximity-induced superconductivity \cite{proximityswnt} in individual single-wall nanotubes has been also demonstrated making these microtubules ideal candidates to realize topological frustrated systems.\\
Interestingly, carbon nanotubes have been proposed as the active element of chemical sensors \cite{chemsensor} since environmental molecules can create chemical bonds randomly distributed along the nanotube. When a U-shaped track with a narrow gauge is formed by a nanotube, an environmental molecule can be captured creating a chemical bond between the rails. Under this condition, the system geometry is equivalent to the case of a legged-ring. Moreover, the captured molecule can move along the system under the combined effect of random and determinist forces, the latter being originated by the tendency to minimize the system free energy. Molecule influence can be modeled as an effective movable hopping between distant sites whose equilibrium position is determined by the initial condition and by the position of local minima of free energy. Since the total energy of the system is affected by the bond position, the question arises whether the local minima of free energy correspond or not to topological non-trivial configurations.\\
Inspired by these questions, we study the topological properties of a Kiteav chain in the shape of a legged-ring, which is here referred to as Kitaev tie. A Kitaev tie with a single movable bond (i.e. the tie knot) can be considered as an oversimplified model to describe the nanotube-based topological device described above (see Figure \ref{model}).\\
The paper is organized as follows. In Section \ref{1} we introduce the Kitaev tie model and its Hamiltonian analysing the topological phase diagram as a function of the movable bond position and the chemical potential. Topological phase diagram structures are discussed in connection with the topological frustration of the system. Appearance of trivial and non-trivial phases is further characterized in terms of the local Majorana polarization. In Section \ref{2} the equilibrium configurations of the movable bond are analyzed by studying the free energy of the system. Conclusions are given in Section \ref{3}.

\section{Model}
\label{1}
\begin{figure}[h!]
	\centering
	\includegraphics[scale=0.85]{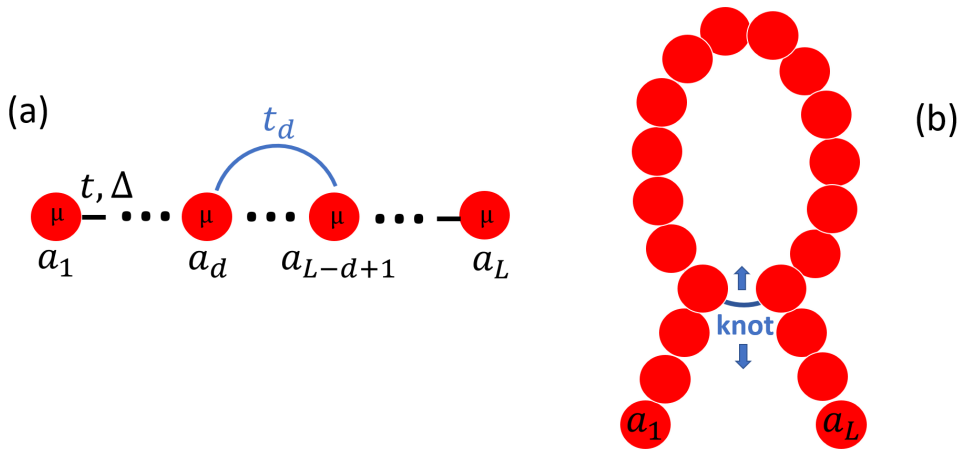}
	\caption{(a) Equivalent tight-binding model of a Kitaev tie system, shown in (b).}
	\label{model}
\end{figure}

\begin{figure}[h!]
 	\centering
 	\includegraphics[scale=0.6]{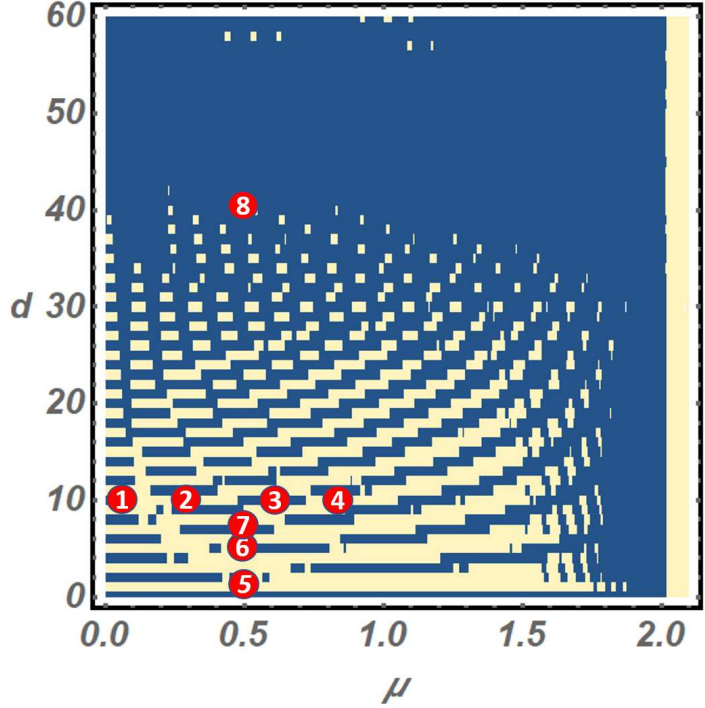}
 	\caption{Topological phase diagram of a Kitaev tie ($L=121$). Blue (white) regions represent topological (non-topological) phases. The model parameters have been fixed as: $t=t_d=1$,  $\Delta=0.02$.}
 	\label{phasediagram}
 \end{figure}

\begin{figure}[h!]
	\centering
	\includegraphics[scale=0.8]{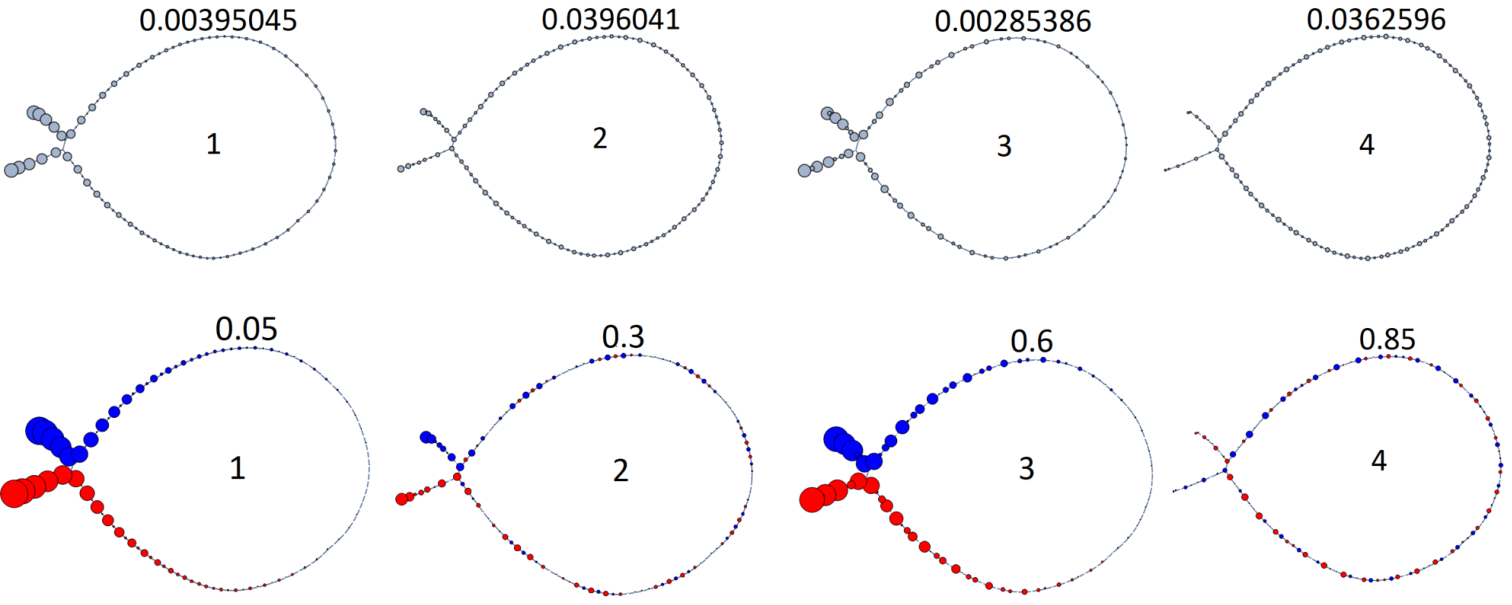}
	\caption{Kitaev tie systems ($L=121$) with knot position $d=10$. Labels $1,\ldots,4$ correspond to the selected points in Figure \ref{phasediagram} and are associated to chemical potential values $\mu=0.05$, $0.3$, $0.6$, $0.85$, respectively. Upper graphs represent lowest energy eigenstates, being the circles diameter proportional to the modulus squared of the wave functions. As for the lower graphs, the circles diameter is proportional to the magnitude of local Majorana polarization. Blue and red colors identify positive and negative polarization values, respectively. Energy eigenvalues associated to the eigenstates (upper graphs) and the values of $\mu$ (lower graphs) are reported in vicinity of each graph. The remaining model parameters have been fixed as: $t=t_d=1$,  $\Delta=0.02$.}
	\label{orizzontalCut}
\end{figure}

\begin{figure}[h!]
	\centering
	\includegraphics[scale=0.8]{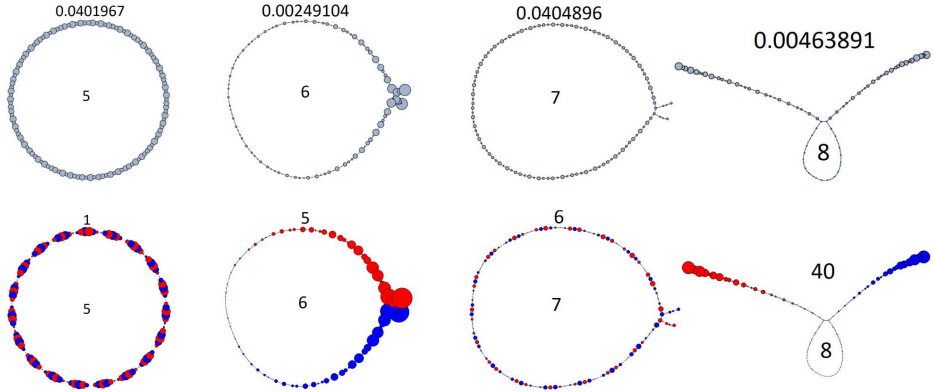}
	\caption{Kitaev tie systems ($L=121$) with fixed chemical potential $\mu=0.5$. Labels inside the graphs correspond to the selected points in Figure \ref{phasediagram} and identify the four knot coordinates $d=1$, $5$, $6$, $40$, respectively. Energy eigenvalues (upper graphs) and the knot coordinate $d$ (lower graphs) are reported in close vicinity of graphs. The remaining model parameters have been fixed as: $t=t_d=1$,  $\Delta=0.02$.}
	\label{verticalCut}
\end{figure}

We study the topological properties of a Kitaev tie realized by perturbing a Kitaev chain with an extra hopping linking distant sites. In our discussion, distant lattice sites, whose position is labeled by the discrete coordinates $\alpha$ and $\beta$, present curvilinear distance $\mathcal{D}(\alpha,\beta)=|\alpha-\beta|$ greater than one. The tight-binding Hamiltonian of the system can be presented in the form:
\begin{equation}
H=H_K+H_d,
\label{Hamiltonian}
\end{equation}
where $H_K$ is the Hamiltonian of the isolated Kitaev chain \cite{Kitaev},
\begin{equation}
H_K=\sum_{j=1}^L[-\mu a^\dagger_{j}a_{j}+(\Delta a^\dagger_{j+1}a^\dagger_{j}-ta^\dagger_{j}a_{j+1}+h.c.)],
\end{equation}
while $H_d$ defines the knot Hamiltonian
\begin{equation}
H_d=-t_d( a^\dagger_{d}a_{L-d+1}+h.c.),
\label{KontHamiltonian}
\end{equation}
linking the distant sites labeled by the discrete coordinates $d$ and $L-d+1$. The system Hamiltonian has been written in terms of fermionic annihilation (creation) operators $a_j$ ($a^\dagger_j$), while the quantity $t>0$ represents the hopping amplitude, $\Delta>0$ is the amplitude of the superconducting pairing, $\mu$ represents the chemical potential and $t_d>0$ is the hopping between distant sites. The knot position is controlled by the discrete coordinate $d$ and can be changed to model a legged ring with variable circumference length (see Figure \ref{model}).\\
Introducing the Nambu representation $\Psi=(a_1,a^\dagger_1,...,a_N,a^\dagger_N)^T$, the Hamiltonian (\ref{Hamiltonian}) can be presented in the Bogoliubov-de Gennes form
\begin{equation}
H=\frac{1}{2}\Psi^\dagger H_{BdG} \Psi,
\end{equation}
where $H_{BdG}$ is a $2L\times 2L$ matrix with $L$ being the number of chain sites. The energy eigenvalues of $H_{BdG}$ can be obtained by numerical diagonalization of a finite-size system, which is here set to $L=121$ lattice sites.
\begin{figure}[h!]
	\includegraphics[scale=0.825]{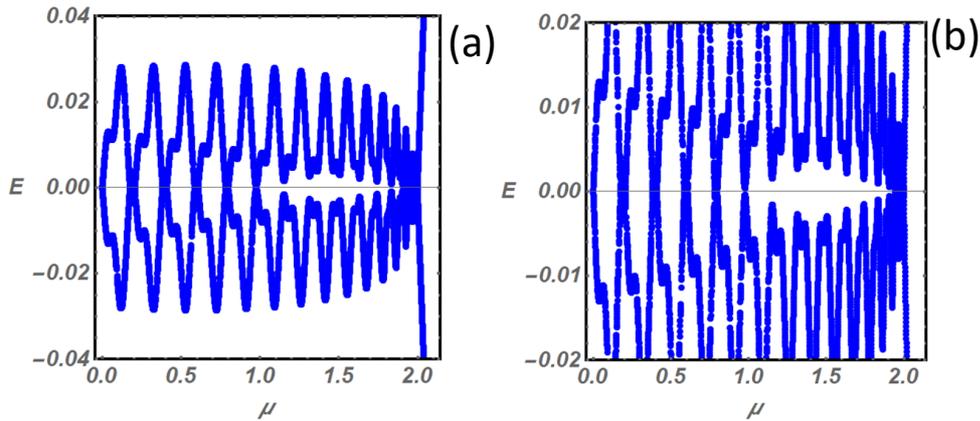}
	\label{E_mu}
	\caption{(a) Low-energy eigenvalues of a Kitaev tie system with $L=121$ lattice sites. The model parameters have been fixed as: $t_d=t=1$, $\Delta=0.02$, $d=30$. (b) Magnified version of panel (a) evidencing the interstitial nature of the topological phase.}
\end{figure}
The results of this analysis are presented in Figure \ref{E_mu} where the low-energy part of the energy spectrum is shown. In particular, the lowest energy eigenvalues are presented as a function of the chemical potential $\mu$ by fixing the knot position to the coordinate $d=30$. The remaining model parameters have been fixed inside the topological non-trivial region of the unperturbed Kitaev chain. The analysis of Figure \ref{E_mu} for $\mu<2t$ clearly shows that there exist optimal chemical potential values for which a topological phase is stabilized. This behavior emerges as a topological frustration effect in which the system reaches a compromise between the topological non-trivial phase of the open Kitaev chain and the trivial behavior of a Kitaev chain with periodic boundary conditions. In balancing these opposite tendencies, topological non-trivial phase develops an interstitial character and nucleates inside a trivial-phase region only when special chemical potential values are selected.\\
In order to get a complete view of the topological properties, in Figure \ref{phasediagram} we show the topological phase diagram in the $\mu$-$d$ plane, where white regions (blue regions) are associated to non-topological (topological) phases. The analysis of the topological phase diagram clearly shows that non-trivial phases (blue regions) nucleate inside trivial ones (white regions) for $\mu \leq 1.8 t$ and $d \le 40$. This behavior is quite expected since, under the considered conditions, the system is similar to a non-topological Kitaev ring perturbed by two short legs. Under appropriate conditions dictated by interference effects, the legs influence can drive the system into a topological phase which emerges as an interstitial phase. The trivial phase disappears when the circumference of the ring is reduced (i.e. for $d>40$), i.e. when the system goes towards the perturbed-chain limit, or when the chemical potential approaches the value $2t$.\\
The topological phase diagram shown in Figure \ref{phasediagram} has been obtained by comparing the energy eigenvalues closer to zero for the Kitaev tie system ($t_d \neq 0$) and for a Kitaev chain ($t_d=0$) of the same size.\\
In order to obtain a further characterization of the topological phase, we use the Majorana order parameter introduced in Ref. \cite{PascalSimon}. Accordingly, the local Majorana polarization can be written as,
 \begin{equation}
 P_M(\omega,n)=2 \sum_{m} \delta(\omega-\epsilon_m) u^{(m)\ast}_n v^{(m)}_n
 \label{MajoranaPolarizaion}
 \end{equation}
 where $u^{(m)}_n$ ($v^{(m)}_n$) is the electron (hole) component of the $m-th$ eigenstate of the system. As discussed in Ref.\cite{Perfetto}, the Majorana polarization $P_M(\omega)=\sum_{n=1}^{L/2} P_M(\omega, n)$ is a good order parameter to characterize the Majorana nature of a given quantum state because it is zero for trivial states and $\pm 1$ for genuine Majorana states.
We have studied the local Majorana polarization and the modulus squared of the eigenstate with energy eigenvalue closer to zero by setting the model parameters in specific points of the phase diagram (see the numbered red circles inside the phase diagram shown in Figure \ref{phasediagram}).\\
In particular, Figure \ref{orizzontalCut} shows a topological/non-topological phases sequence as the chemical potential $\mu$ is increased at fixed knot position $d=10$. A similar sequence is obtained in Figure \ref{verticalCut} where the discrete knot coordinate $d$ has been increased with fixed chemical potential $\mu=0.5$.
In Figures \ref{orizzontalCut} and \ref{verticalCut} the lattice sites of the Kitaev tie present a radius proportional to the magnitude of the wavefunction of the lowest energy eigenstate (upper row) or to the local Majorana polarization (lower row). In the latter case, blue and red circles identify positive and negative values of the Majorana polarization, respectively. The analysis of the mentioned figures directly shows that topological states correspond to Majorana modes localized on the legs of the system, while presenting a decaying behavior inside the ring. On the other hand, trivial states are uniformly distributed throughout the system with a zero Majorana polarization. The case of the Kitaev ring, which is a trivial system, is shown in Figure $5$ first column for comparison.

\label{Section2}
\section{Formation mechanism of a Kitaev tie configuration}
\label{2}
\begin{figure}[h!]
	\centering
	\includegraphics[scale=0.8]{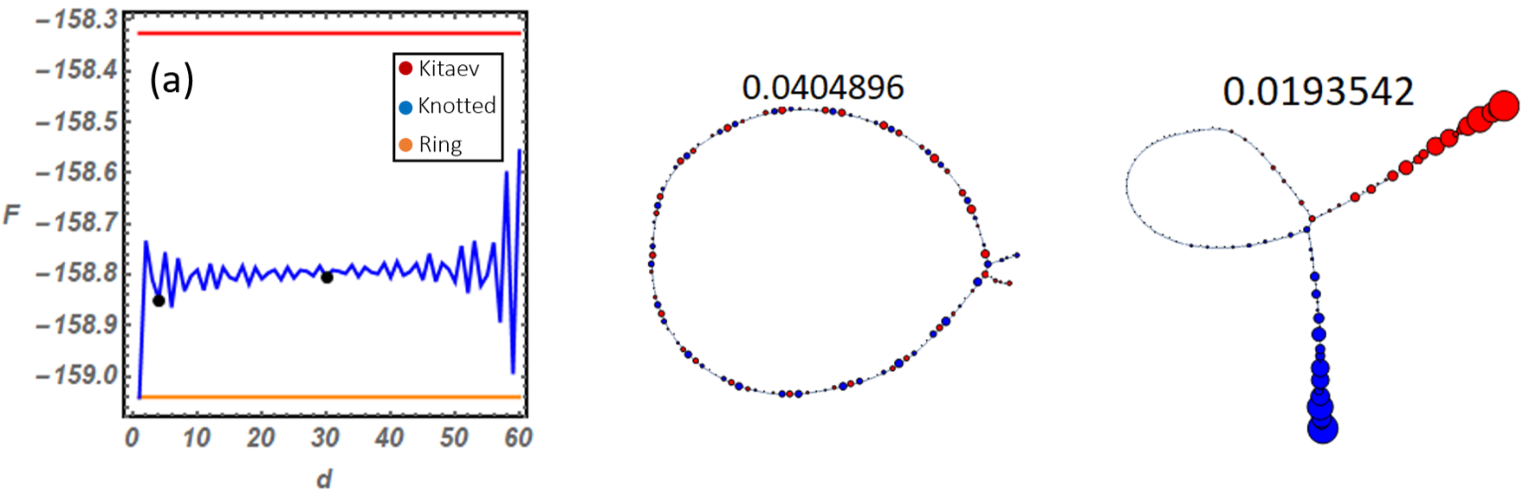}\\
	\includegraphics[scale=0.8]{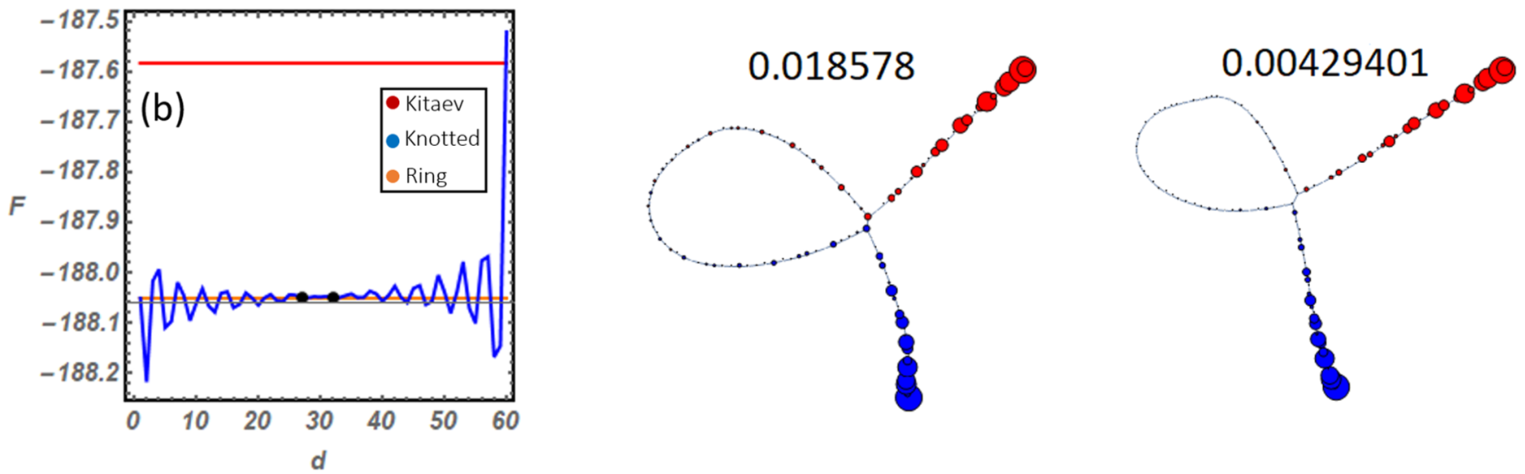}\\
	\includegraphics[scale=0.8]{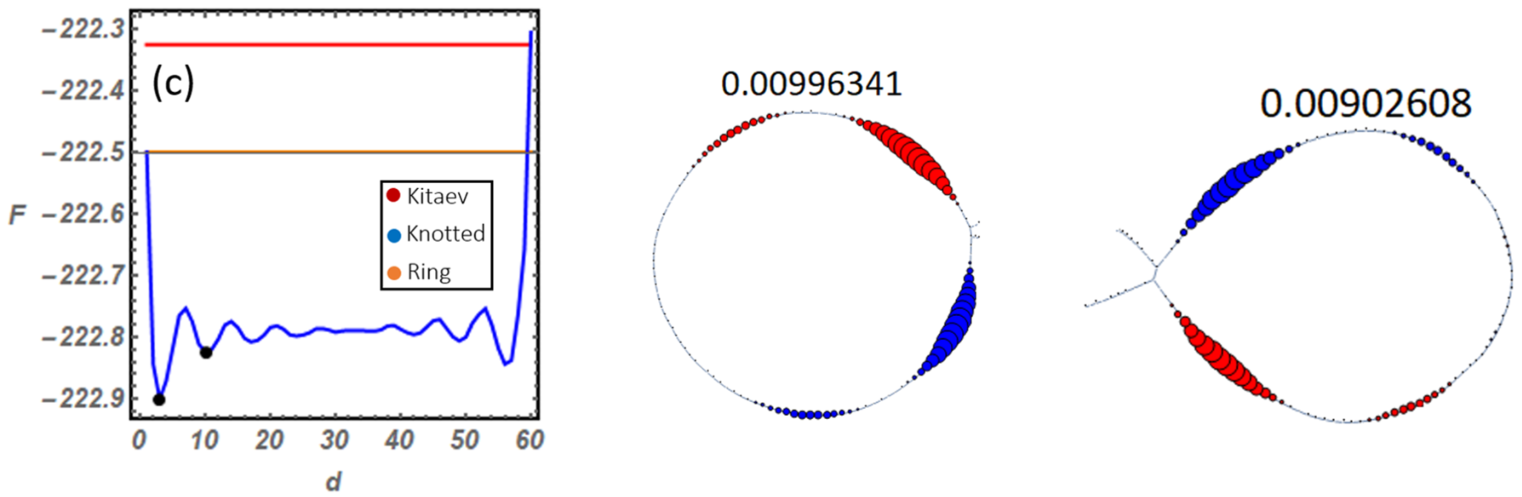}
	\caption{(a) Free energy of a Kitaev tie system as a function of the knot coordinate $d$ (blue curve) for $\mu$ = $0.5$. Free energy of a Kitaev ring (orange curve) and a Kitaev chain (red curve) are reported for comparison. Side graphs represent the local Majorana polarizations for system realizations which correspond to black dots ($d=4$, $d=30$) on the free energy curve. Energy eigenvalues corresponding to the eigenmodes are reported in close vicinity of the graphs. Chemical potential has been fixed to $\mu=1.3$ ($\mu=1.8$) in panel (b) (panel (c)). The model parameters have been fixed as: $t=t_d=1$, $\Delta=0.02$, $k_B T=\Delta/10$ and $L=121$. }
	\label{F_KRK}
\end{figure}

\begin{figure}
	\includegraphics[scale=0.6]{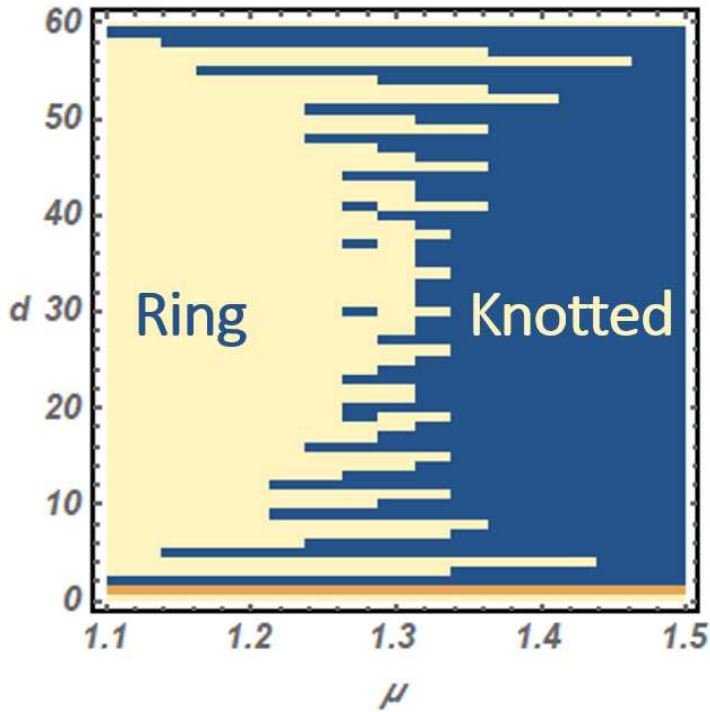}
	\caption{Topological stability diagram representing the stability regions in $\mu$-$d$ pane of a Kitaev tie (knotted phase) and of a Kitaev ring (ring phase). The model parameters have been fixed as: $t=t_d=1$,  $\Delta=0.02$, $k_B T=\Delta/10$, while the system size has been fixed to $L=121$ lattice sites.}
	\label{sgn(FKn-FR)}
\end{figure}

In this section we explore the formation mechanism of a Kitaev tie starting from a curved Kitaev chain. In particular the question arises wether a Kitaev chain with open boundary conditions is thermodynamically stable against the spontaneous or assisted formation of an extra bond between distant sites. Indeed, when a Kitaev chain is bent to form a U-shaped conductor, distant lattice sites (from the curvilinear distance point of view) can create extra bonds, which are not permitted in a straight Kitaev chain. The latter situation can be realized in carbon nanotubes which are flexible ballistic conductors. In order to follow our program, we consider the free energy of the system within the BdG formalism \cite{FreeEnergy}:
\begin{equation}
F=-2k_BT\sum_i \ln \biggl( 2 \cosh \frac{E_i}{2k_BT}\biggr)
\label{FEquation}
\end{equation}
which can be written in terms of positive eigenvalues $E_i$ of $H_{BdG}$, while $k_B$ represents the Boltzmann constant and $T$ the absolute temperature. The information on the geometric configuration of the system (i.e. Kitaev ring, legged-ring or Kitaev chain) is stored within the single-particle spectrum which depends on the boundary conditions imposed by the hopping matrix and on the chemical potential. In real systems chemical potential can be changed by using electrostatic gates as routinely done in field effect nanodevices \cite{fetcnt}. Thus, we consider the chemical potential as a free parameter being it variable by using external electrostatic potentials. For the sake of simplicity we disregard multiple-knot configurations (e.g. $8$-shaped systems) whose occurrence, under specific conditions, can be excluded.\\
Resorting to numerical analysis, we have studied the free energy of a Kitaev tie (Eq. \ref{FEquation}) as a function of knot position, controlled by the discrete coordinate $d$, at fixed values of the chemical potential $\mu$. Thermal energy has been fixed to $k_B T=\Delta/10$, which is compatible with a stable superconducting phase, the latter being a necessary condition to study topological properties.\\
The free energy analysis is reported in Figure \ref{F_KRK}. As a general comment we do observe that a Kitaev chain, under appropriate curvature conditions, is unstable against the formation of an extra bond. This is clearly visible observing that the Kitaev chain free energy takes higher values compared to those of a Kitaev ring and a Kitaev tie, being the latter more stable configurations towards which the system relaxes. The arrival configuration, i.e. a Kitaev ring or tie, depends on the initial conditions of the system. The free energy of a Kitaev tie presents a complicate oscillating behaviour as a function of the discrete coordinate $d$ defining the knot position. The average value of free energy of a Kitaev tie depends on the chemical potential and can take intermediate or lower values with respect to those of a Kitaev chain and a Kitaev ring. Setting the chemical potential value to $\mu=1.3$, a global minimum of the free energy is reached alternatively for a Kitaev ring or a Kitaev tie depending on the position of the tie knot. This behavior highlights the frustrated nature of the system.\\
When the system is shaped in the form of a Kitaev tie and a relaxation towards a Kitaev ring is not possible, the knot position is decided by the initial conditions and by the local minima of the free energy, which are stable configurations for the system. A local minimum is reached under the action of a restoring force $\mathcal{F} \approx - \partial_{d}F(d,\mu)$ determined by the free energy gradient, where $F(d,\mu)$ plays the role of an effective potential for the position of the extra bond.
Once a local minimum has been reached, thermal fluctuations can in principle activate transitions between adjacent minima. According to the Arrhenius rate equation, these transitions are more frequent when thermal energy $k_B T$ is greater than the free energy barrier $\Delta F$ separating two system configurations. Direct inspection of the Kiteav tie free energy shows that local minima are in general sufficiently deep to prevent transitions between adjacent minima. However, when free energy is sufficiently flat with respect to a variation of the knot position, stochastic fluctuations can change the tie knot position. This condition is for instance verified in Figure \ref{F_KRK} (b), where the system free energy around $d \approx 30$ is almost flat and the extra bond can freely move under the action of stochastic fluctuations. Stochastic dynamics of the extra bond can be studied by using Langevin or master equation approach complemented by the reasonable assumption that the time evolution of the mechanical degree of freedom (i.e., the knot position) is adiabatic in comparison with the typical time scales on which the electronic part of the problem reacts to external perturbations. The analysis of the stochastic motion of the extra bond, which is however beyond the purposes of this work, can provide relevant clues for the development of proof-of-principle nanomachines \cite{nanomotors} working as topological nanomotors.\\
The analysis of the site-dependent Majorana polarization (see Figure \ref{F_KRK}) shows that free energy minima of a Kitaev tie not necessarily correspond to topological non-trivial configurations.\\
When the chemical potential is set to $\mu=1.8$, i.e. close to the phase boundary of the unperturbed Kitaev chain, the knotted configuration corresponding to a Kitaev tie system is thermodynamically favored. Under this condition, starting from a Kitaev ring, the system relaxes towards a kitaev tie, as shown in  Figure \ref{F_KRK} (c). Interestingly, we have demonstrated that in close vicinity of the phase boundary of the unperturbed Kitaev chain a topological trivial Kitaev ring relaxes into a topological non-trivial phase under the effect of an effective potential acting on the extra bond. The mentioned phenomenon evidences that, under appropriate conditions, trivial systems manifest a spontaneous tendency to establish a topological non-trivial phase.\\
The competition between the stable system configurations, i.e. the Kitaev ring and the Kitaev tie, is studied in Figure \ref{sgn(FKn-FR)}. The phase plane evidences that for $\mu \lesssim 1.2$ ($\mu \gtrsim 1.4$) a Kitaev ring configuration (a kitaev tie configuration) is thermodynamically stable. A strong competition between topological and trivial phase is instead evident inside the transition region $1.2 \lesssim \mu \lesssim 1.4$, where the stable phase is decided by the knot coordinate $d$.\\
As a closing remark, we do observe that interference effects play a relevant role and affect the frustration mechanism involved in the topological phase transition. The latter statement can be easily proven by using a transfer matrix analysis \cite{transferm1,transferm2} adapted to the Kitaev tie system.
\section{Conclusions}
\label{3}
In conclusion, we have shown that a topological frustrated system can be obtained by perturbing a Kitaev chain by introducing an extra bond between distant lattice sites. In real systems this configuration, which is here referred to as Kitaev tie, can be obtained by using flexible ballistic conductors such as the carbon nanotubes. The Kitaev tie shows a rich phase diagram as a function of the chemical potential and of the knot position, with interstitial non-trivial phases immersed inside a trivial region. A thermodynamic stability analysis evidences that a curved Kitaev chain is unstable against the spontaneous formation of an extra bond between distant lattice sites and relaxes, if possible, towards a trivial Kitaev ring or a Kitaev tie, whose topological properties depend on the knot position. Moreover, by setting the chemical potential in close vicinity of the phase boundary of an unperturbed Kitaev chain (i.e. $\mu = 1.8$), we have demonstrated that a trivial Kitaev ring relaxes towards a Kitaev tie, with topological non-trivial properties. The above observation leads to the conclusion that under appropriate circumstances topological trivial systems can manifest the spontaneous tendency to establish a topological order. These findings could be of interest for the emerging field of quantum biology \cite{qbio} in view of the possible answer to the question wether topological properties play or not some role in determining selected aspects of the peculiar behavior of living matter.

\end{document}